\documentclass[prl,twocolumn,showpacs,floatfix]{revtex4}
\usepackage[T1]{fontenc} 
\usepackage[cp1250]{inputenc}
\usepackage{graphicx} 
\usepackage[english]{babel}

\newcommand{\bq}{\begin{equation}} 
\newcommand{\eq}{\end{equation}}
\newcommand{\ba}{\begin{eqnarray}} 
\newcommand{\ea}{\end{eqnarray}}

\begin{document}
\title{Mutational pathway determines whether drug gradients accelerate evolution of drug-resistant cells}
\author{Philip Greulich$^*$}
\author{Bart{\l}omiej Waclaw}
\thanks{Equal contribution}
\author{Rosalind J. Allen}
\affiliation{SUPA, School of Physics and Astronomy, University of Edinburgh, Mayfield Road, Edinburgh EH9 3JZ, United Kingdom}
\noindent
\begin{abstract}
Drug gradients are believed to play an important role in the evolution of bacteria resistant to antibiotics and tumors resistant to anti-cancer drugs. 
We use a statistical physics model to study the evolution of a population of malignant cells exposed to drug gradients, where drug resistance emerges via a mutational pathway involving multiple mutations. We show that a non-uniform drug distribution has the potential to accelerate the emergence of resistance when the mutational pathway involves a long sequence of mutants with increasing resistance, but if the pathway is short or crosses a fitness valley, the evolution of resistance may actually be slowed down by drug gradients. These predictions can be verified experimentally, and may help to improve strategies for combatting the emergence of resistance. 
\end{abstract}
\maketitle

The evolution of drug resistance is an urgent problem in the treatment of disease, from bacterial infections to cancer. Attempts to address this problem include  the characterization of  mutational pathways leading to resistance \cite{weinreich, toprak-kishony}, as well as theoretical \cite{this_1960,bonhoeffer,djaustin,lipsitch,wang,bergstrom} and experimental \cite{multidrug,torella, jumbe} studies of the emergence of resistance under different treatment regimens. These studies usually assume a spatially uniform drug concentration. However, in many clinical situations  drug concentrations vary in space \cite{clin1,clin2}, for example where malignant cells form less drug-permeable layers such as bacterial biofilms \cite{stewart} or tumour stromas \cite{tredan}. Recent experimental work \cite{austin} suggests that the evolution of antibiotic resistance in bacterial populations can be greatly accelerated if the antibiotic concentration is spatially non-uniform. 

It is often observed that several mutations are required to obtain maximal resistance to a drug \cite{weinreich,toprak-kishony,austin}. In some cases, fitness (i.e.~drug resistance) increases steadily along the mutational pathway to full resistance \cite{weinreich}; in other cases,  epistatic interactions between mutations may result in less fit intermediate genotypes (fitness ``valleys'') \cite{schrag1, schrag2, weinreichrev}. The role of mutational pathways in controlling evolutionary dynamics has been studied in the quasispecies model \cite{quasispecies,krug} and in models of cancer progression \cite{antal1,antal2}. That work does not take into account the effects of spatial inhomogeneity. In models without complex evolutionary pathways, it is well known that  spatial structure can increase genetic diversity, the rate of evolutionary diversification \cite{gastner, kirkpatrick, barton, pekalski, meszena}, and the rate of viral drug resistance \cite{kepler}; indeed, in a broader statistical physics context, spatial structure plays a key role in many theoretical studies of evolving populations \cite{gastner, frey1,martens,martens2,ali,mobilia,BW,HH}.

\begin{figure}
\includegraphics*[width=\columnwidth]{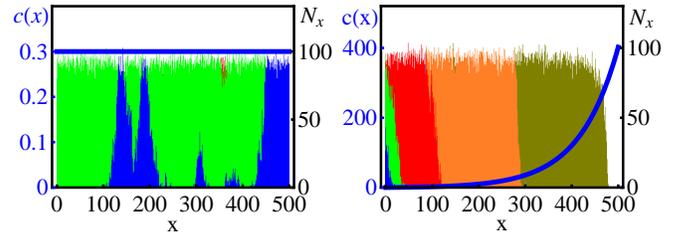}
\caption{\label{fig:example} Simulation snapshots for the cases of a uniform and an exponentially increasing drug concentration (left and right, respectively). Blue lines show the drug concentration (left axes), while the colours represent the populations of the different genotypes (right axes). Parameter values are $K=100,L=500,M=6,\mu=5\times 10^{-6},\beta_m=4^{m-1}$, and the drug concentration $c=0.3$ (left panel) and $c(x)=e^{\alpha x}-1$ with $\alpha=0.012$ (right panel). For corresponding movies see \cite{suppl0}.}
\end{figure}

Here, we present a model that combines evolution along mutational pathways in genotype space with population dynamics in non-uniform environments. We use this model to study the evolution of drug resistance as a population of cells colonizes an environment containing a non-uniform  distribution of a growth-inhibiting drug. Our key result is that the effect of drug gradients depends critically on the pathway to resistance. This is due to a complex interplay between the spatial drug distribution and the mutational pathway. In the presence of drug gradients, the population evolves drug resistance in a sequence of waves of increasingly better adapted mutants that extend its range in a step-wise manner. In contrast, for a uniformly distributed drug, resistant mutants evolve at random positions and spread over the entire environment. If tolerance to the drug increases monotonously along the mutational pathway, drug gradients can significantly accelerate the evolution of resistance by increasing selection at the population's edge. However, if the pathway crosses a fitness valley, evolution of resistance may actually be slowed down by a non-uniform drug distribution, as a result of a reduced rate of ``stochastic tunnelling'' due to a smaller population size.

{\it The model.}
We consider a growing population of cells which mutate between $M$ possible genotypes, with different levels of resistance to a drug. To model the effects of spatial heterogeneity, the population is assumed to reside within a chain of $L$ connected microhabitats, which may contain different concentrations of the drug. These discrete microhabitats might represent connected chambers in a microfluidic experiment \cite{austin}; in the limit of small microhabitats, our model represents a population growing in continuous space. Within a given microhabitat $i$ the population is assumed to be well-mixed, with a fixed carrying capacity $K$; cells of genotype $m$ replicate at rate $\phi_m(c_i)(1-N_{i}/K)$ where $N_i$ is the total population of cells in microhabitat $i$,  $c_i$ is the drug concentration and $\phi_m(c_i)$ is the growth rate of genotype $m$, which depends on the local drug concentration. Upon replication, cells mutate with probability $\mu$; we consider the case of an unbranched mutational pathway to drug resistance, such that genotype $m$ mutates only into genotypes $m \pm 1$ (without any bias). Cells migrate between microhabitats $i$ and $i \pm 1$ at rate $b/2$ and die at a fixed rate $d$; the latter ensures that in the steady state there is a turnover of cells, with the net birth rate being equal to the death rate $d$.

A key feature of our model is the fact that different genotypes show different levels of drug resistance: genotype 1 is least resistant while genotype M is most resistant. For all genotypes, the growth rate decreases with the drug concentration; the \emph{minimal inhibitory concentration} (MIC), $\beta_m$, denotes the drug concentration at which genotype $m$ ceases to be able to grow. This is embodied in our model by setting $\phi_m(c) = \max\left\{0,1-\left({c}/{\beta_m}\right)^2\right\}$. This choice is inspired by Ref. \cite{regoes} (see also the supplementary text \cite{suppl}, Sec. I).

We study this model using kinetic Monte Carlo simulations (\cite{suppl}, Sec. II) which are initiated with $N_1=K$ cells of genotype 1 in microhabitat 1 and with all the other microhabitats empty, so that the population colonizes the space during the simulation, as it evolves resistance to the drug. We define the units of time by fixing the maximal growth rate $\phi_m(0) = 1$ and the units of drug concentration by fixing $\beta_1=1$, and we set  $b=0.1$, $d=0.1$, and $K=100$. We use values of $\mu$ and $L$ such that the number of mutants per generation emerging anywhere in the environment is typically small, $\mu K L \leq 1$, and stochastic effects are relevant.  A detailed discussion of the parameter set is given in the supplementary text \cite{suppl}, Sec. I.

To investigate the effects of the spatial distribution of the drug, we consider two scenarios: i) a non-uniform drug concentration $c_i=\exp(\alpha i)-1$, which increases exponentially from left to right with steepness $\alpha$, and ii) a uniform drug concentration $c_i \equiv c$. Note that we have not chosen to keep the total amount of drug constant. Rather, we allow the parameters $\alpha$ and $c$ to vary over their whole range, and determine whether, {\em{under any circumstances}} (i.e. for any values of $\alpha$ and $c$), it is possible for resistance to evolve faster in the non-uniform environment. In the non-uniform case, we vary $\alpha$, the only constraint being that the drug concentration $c_1$ in the first microhabitat must be lower than $\beta_1$, to allow the first genotype to establish. In the uniform case, we vary the drug concentration $c$; the same condition implies that $c\leq \beta_1$, otherwise the space cannot be colonized and no mutants can emerge. Obviously, the local drug concentration can be much higher in the non-uniform than the uniform environment. Indeed, one of the results of our model is that spatial gradients can allow colonization of regions of much higher drug concentration than would be possible in a uniform drug environment. 

\begin{figure}
\includegraphics*[width=\columnwidth]{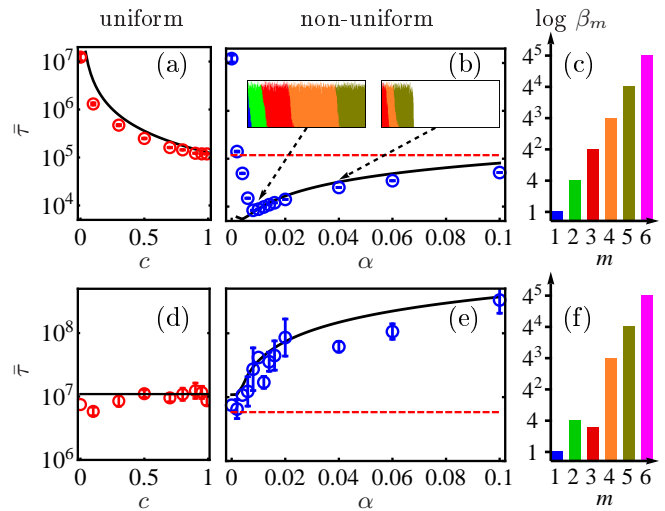}
\caption{\label{fig:times_comp} Average time to resistance $\bar \tau$ for uniform ((a,d), red circles) and non-uniform ((b,e), blue circles) drug concentrations, for $M=6,L=500, K=100$, and $\mu=5\times 10^{-6}$. Panels (a,b): exponentially increasing MIC (shown in c), panels (d,e): fitness valley (shown in f). For the non-uniform case (b,e),  the red dashed lines show the minimal value of $\bar \tau$ obtained for the uniform case (i.e. the minimum of $\bar \tau(c)$ from (a,d)). The black lines show the theoretical predictions: (a): $\bar\tau \approx 126642/c^{3/2}$ (see also Eq.~(IV.11) from Supp. Mat. \cite{suppl}, Sec. IV), (b):  Eq.~(\ref{eq:pred}), (d): Eq.~(\ref{eq:tau_vh}), and (e): Eq.~(\ref{eq:tau_v}). The insets show simulation snapshots taken just before the first occurrence of genotype $m=6$, for two values of $\alpha$ (indicated by arrows).}
\end{figure}

{\it Monotonically increasing MIC.}
The mutational pathway to drug resistance plays a crucial role in our simulations. We first consider a pathway to resistance for which the $M$ genotypes have increasing levels of drug resistance -- i.e. $\beta_m >  \beta_{m-1}$ for all $m>1$, as depicted in Fig.~\ref{fig:times_comp}c. In particular, we set $M=6$ and $\beta_m=4^{m-1}$; the ratio $\beta_6/\beta_1 \approx 10^3$ between fully resistant and wild-type cells is consistent with experimentally determined values \cite{weinreich}. 

Our simulations show that the emergence of drug resistance occurs very differently in the cases of uniform and non-uniform drug distribution, as illustrated in the  snapshots of Fig.~\ref{fig:example} and in the supplementary movies \cite{suppl0}. If the drug concentration is spatially uniform  (Fig.~\ref{fig:example} left), genotype 1 (blue) first spreads to fill the entire space, then mutants of genotype 2 (green) emerge at random locations and spread to fill the space (competing with cells of genotype 1) before giving rise to more resistant mutants of genotype 3, etc. In contrast, in the presence of a drug gradient (Fig.~\ref{fig:example} right), population waves of increasingly better-adapted mutants advance from left to right in a step-wise manner.   Genotype $m$  colonizes the space only up to a well-defined  spatial boundary, where it forms a stationary ``front''; better-adapted mutants then emerge from this front to further colonize the space. Thus the spatial gradient generates local ``niches'' with low drug concentrations, in which less resistant genotypes can dwell,  generating more resistant mutants, which can then colonize regions with higher drug concentrations.

These differences have important consequences for the time to emergence of drug resistance. Fig.~\ref{fig:times_comp}a and b shows the mean time $\bar \tau$ to emergence of full drug resistance -- i.e.  the time to emergence of a mutant with $m=M=6$, averaged over surviving populations.  For a uniform drug concentration $c_i=c$, $\bar \tau$ decreases as the drug concentration $c$ increases (note that $c=1$ corresponds to the MIC of genotype 1) (Fig.~\ref{fig:times_comp}a). For the non-uniform drug distribution (Fig.~\ref{fig:times_comp}b), $\bar \tau$ varies non-monotonically as a function of the steepness $\alpha$, with a minimum at $\alpha \approx 0.01$. This arises because for very small $\alpha$ the selection pressure for the evolution of resistant mutants is low (since little drug is present), whereas for very large $\alpha$, the fronts become narrow, reducing the size of the ``zone'' in which new resistant mutants can emerge (see snapshots in Fig.~\ref{fig:times_comp}b).

Importantly, for almost all values of $\alpha$ and $c$ in Fig.~\ref{fig:times_comp}, resistance emerges faster for the non-uniform drug distribution than for the uniform case; the minimal value of $\tau(\alpha)$ in the non-uniform case is smaller by an order of magnitude than the minimal value of $\tau(c)$ in the uniform case (dashed line in Fig.~\ref{fig:times_comp}b).  This can be understood intuitively as follows. For the uniform drug distribution, a new genotype $m+1$ must compete with the already established genotype $m$; selection pressure is weak because the drug concentration is low (restricted by the constraint $c<\beta_1$). In contrast,  the non-uniform drug distribution ensures that for each genotype $m$, there exists a spatial location with drug concentration $c$ close to its MIC $\beta_m>\beta_1$. At the population's front, the high drug concentration imposes maximal selection pressure for the emergence of the next genotype $m+1$, which then can colonize the adjacent empty space, free from competition. Thus, if the MIC increases monotonically along the pathway to resistance, a non-uniform drug distribution carries the potential for much faster evolution of drug resistance than is possible if the drug is uniformly distributed.

\begin{figure}
\includegraphics*[width=0.6\columnwidth]{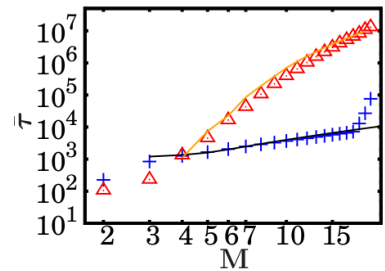}
\caption{\label{tau_M_fig} Average time $\bar \tau$ to full resistance as a function of the mutational pathway length $M$ for uniform ($c=0.9$; triangles) and non-uniform ($\alpha=0.07$; crosses) drug distribution. In both cases $L=300, K=100$, and $ \mu=10^{-4}$. Solid lines are theoretical predictions for the non-uniform case (black line, calculated numerically from Eqs. (\ref{eq:T_mut}) and (\ref{eq:pred})) and uniform case (calculated as explained in the supplementary text \cite{suppl}, Sec. V).}
\end{figure}

This picture depends crucially, however, on the  length of the mutational pathway, as shown in Fig.~\ref{tau_M_fig}. For long mutational pathways (large $M$), resistance indeed emerges faster in the non-uniform environment. For short pathways ($M<4$ in our simulations), however, the situation is reversed; resistance  actually emerges faster in the uniform environment than in the non-uniform one, despite the fact that the drug concentration is higher in the latter case. This is because  the uniform environment provides a larger population size from which resistant mutants can emerge. 

Our results can be rationalized using simple physical arguments. We begin with the case of a non-uniform drug distribution. In our simulations, the population wave of genotype $m$ typically reaches a "steady state" before  mutants of genotype $m+1$ emerge; this is because the selection pressure is high only at the stationary front (see the supplementary movies \cite{suppl0}). We therefore consider these two processes separately. In the continuous approximation (valid for large $L$ and $\alpha\ll 1$), the expansion of a wave of mutants of genotype $m$ is described by the Fisher-KPP equation \cite{math-biol}:
\begin{eqnarray}
	\partial_t N_m &=& \frac{b}{2} \partial_{xx} N_m + \phi_m N_m \left(1-\frac{N_m}{K}\right)-dN_m \label{eq:fisher}   \\
	&=& \frac{b}{2} \partial_{xx} N_m + (\phi_m-d) N_m \left[1-\frac{\phi_m N_m}{K(\phi_m-d)}\right] \nonumber \,\,\, ,
\end{eqnarray}
where $x\equiv i$, $\phi_m\equiv \phi_m(c(x))$ and $N_m\equiv N_m(x,t)$ denotes the population of genotype $m$.  If $c(x)\ll \beta_m$, $\phi_m \approx 1$ and  Eq.~(\ref{eq:fisher}) describes a Fisher wave  propagating with speed $v\approx \sqrt{2b(1-d)}$  \cite{math-biol}. The wave stops when it reaches the point where $c(x)\approx \beta_m$; for small $b$ the stationary solution of Eq.~(\ref{eq:fisher}) reads $N^*_m(x) = K [1 - d/\phi_m(c(x))]$, which decays to zero at $x^*_m =  (1/\alpha)\ln (\beta_m\sqrt{(1-d)}+1)$. Assuming that the wave of mutants of genotype $m$ emerges at $x^*_{m-1}$ (i.e. at the stationary front of the preceding wave),  the time it takes to reach its stationary state is then $T^{\mathrm{wave}}_m\approx (x^*_m-x^*_{m-1})/v$, with $T^{\mathrm{wave}}_1\approx x^*_1/v$. 

Once the stationary population of genotype $m$ is established, the waiting time before a new wave of mutants of genotype $m+1$ arises can be expressed for low mutation rates as the inverse of the total rate at which mutants establish in the population -- i.e., assuming strong selection: 
\begin{equation}
	\label{eq:T_mut}
	T^{\mathrm{mut}}_{m+1}=\left[\frac{\mu}{2}\int_{x^*_{m-1}}^{x^*_m}  N^*_m(x) r(x) P_{\mathrm{fix}}(x) dx \right]^{-1} \, .
\end{equation}
Here $\mu/2$ is the probability to mutate from genotype $m$ to $m+1$, $r(x) \approx d$ is the rate of reproduction in the steady state, and $P_{\mathrm{fix}}=(\phi_{m+1}-\phi_m)/\phi_{m+1}$  is the probability of fixation  of genotype $m+1$; this is a standard result \cite{nowak-book}. 

The mean total time until the first cell of genotype $M$ emerges is then
\begin{equation}\label{eq:pred}
	\bar \tau \approx \sum_{m=1}^{M-1} T^{\mathrm{wave}}_m +   \sum_{m=2}^{M-1} T^{\mathrm{mut}}_{m}. 
\end{equation}  
 The value of $\bar\tau$ calculated from Eqs.~(\ref{eq:T_mut}) and (\ref{eq:pred}) is in good quantitative agreement with our simulation results (black line in Fig.~\ref{fig:times_comp}b). 
Equation~(\ref{eq:pred}) decomposes the time to resistance  into the independent contributions of each wave of mutants. For our choice of MICs and drug distribution, $T^{\mathrm{mut}}_m$ and $T^{\mathrm{wave}}_m$ are approximately independent of $m$ for large $m$ (see supplementary text \cite{suppl}, Sec. III). Hence for our model $\bar \tau$ grows linearly with the length of the pathway $M$ as shown in  Fig.~\ref{tau_M_fig}. The scaling changes, however, for very large $M$ (see \cite{suppl}, Sec. III).

If the drug is uniformly distributed, $c(x)=c$, a different argument applies. Here, new genotypes must compete in an already fully colonized space. The time to fixation of genotype $m+1$  scales with the fitness advantage as $(\phi_{m+1} - \phi_m)^{-\gamma}$, where $\gamma>0$ depends on whether mutations are rare or frequent (see supplementary text \cite{suppl}, Sec. IV). Since the drug concentration is fixed, the selective pressure $\phi_{m+1} - \phi_m \propto c^2/4^{2m}$ decreases for successive genotypes. The rate-limiting step in the evolution of resistance is then the fixation of genotype $M-1$, and, as we show in the supplementary text \cite{suppl}, Sec. IV, for the parameter set from Fig.~\ref{fig:times_comp}, we obtain $\bar\tau \approx 126642/c^{3/2}$. This prediction agrees well with our simulation results (Fig.~\ref{fig:times_comp}a). The decrease in selective pressure with $m$ also leads to a super-linear increase in $\bar\tau(M)$ with $M$, as shown in Fig.~\ref{tau_M_fig}. It is a consequence of this effect that, even though for small $M$ evolution is faster in the uniform drug distribution, for large $M$, resistance evolves much faster in the drug gradient \footnote{Note that here we have assumed that the MICs of intermediate genotypes remain fixed as $M$ changes; one can also show (see supplementary text \cite{suppl}, Sec. VI) that the same general conclusions hold if  $\beta_m$ is scaled with $M$ so as to keep the MIC of the most resistant genotype constant. }.

{\it Fitness valley.} We now contrast these results with the situation where the pathway to resistance passes through a "fitness valley" - i.e. one of the intermediate genotypes $m$ has a lower MIC (is less drug-resistant) than its neighbouring genotypes $m-1$ and $m+1$. This scenario can arise due to epistatic interactions between mutations \cite{schrag1,schrag2}, such that two mutations are required to gain a particular fitness benefit. In our model we set $\beta_3=3.5$, keeping all the other $\beta_m=4^{m-1}$ as before, so that $\beta_2>\beta_3<\beta_4$, as depicted in Fig.~\ref{fig:times_comp}f. Figure~\ref{fig:times_comp}d and e shows that the presence of the fitness valley has a dramatic effect on the time to resistance in the non-uniform environment: $\bar \tau(\alpha)$ now rises steeply with $\alpha$. Crucially, the shortest time to resistance in the non-uniform environment is now comparable to that in the uniform environment, $\mathrm{min}_\alpha(\tau(\alpha)) \approx \mathrm{min}_c(\tau(c))$, and $\tau(\alpha)>\mathrm{min}_c(\tau(c))$ for almost all values of $\alpha$. Thus, when the pathway to resistance contains a fitness valley, a non-uniform drug distribution does not speed up, and may well slow down the emergence of resistance.

To understand this result, we argue that the rate-limiting step in the evolutionary process is the ``tunnelling'' through the  fitness valley \cite{chao}: mutants of genotype 4 arise from the population of genotype 2 via short-lived mutants of genotype 3 which do not reach fixation. The tunnelling rate  $\bar{\tau}^{-1}$ has been calculated for well-mixed populations in Ref.~\cite{chao} (Eq.~(2) therein). Applying this result to the case of uniform drug distribution we obtain $\bar{\tau}^{-1} \approx r N_2(\mu/2)^2 (P_{\rm fix}/s)$ where $N_2 \approx L K (1-d)$ is the population size of genotype 2, $s = (\phi_2 -\phi_3)/\phi_2$ is the selective advantage of genotype 2 over genotype 3, $P_{\rm fix} = (\phi_4 -\phi_2)/\phi_2$ is the fixation probability of genotype 4 and $r$ is the growth rate which in the steady state equals the death rate $d$. For our choice of $\phi(c)$ and $\{\beta_m\}$, this gives 
\bq
	\bar{\tau}\approx  1.23/(d\mu^2 N_2) = 1.23/(d\mu^2 L K(1-d)), \label{eq:tau_vh}
\eq
which is independent of $c$. Equation (\ref{eq:tau_vh}) is in good agreement with our simulation results (black line in Fig.~\ref{fig:times_comp}d). Extending this approach to the non-uniform case, we integrate over the steady-state population density of genotype 2:
\bq
	\bar{\tau} \approx \frac{1.23}{d \mu^2}\left[\int_{0}^{x^*_2} N^*_2(x) dx \right]^{-1} \label{eq:tau_v}  .
\eq
This result agrees  well with our simulation results for the non-uniform drug distribution (Fig.~\ref{fig:times_comp}e). The increase in $\bar\tau$ with the steepness of the drug concentration profile $\alpha$ occurs because the domain occupied by genotype 2 decreases as $\alpha$ increases; the non-uniform drug distribution decreases the  steady state population size of genotype 2, reducing the pool of cells from which mutants of genotype 4 can emerge and  slowing down the evolution of resistance. 

{\it Conclusion.} Our results show that the mutational pathway to drug resistance plays a crucial role in determining the effect of a spatial drug distribution on the time to evolve drug-resistant cells. If fitness (i.e. level of drug resistance) increases monotonically along the mutational pathway, a non-uniform drug distribution has the potential to accelerate the evolution of resistance, by a factor that  increases dramatically with the length of the pathway. In contrast, for short pathways, or those involving a fitness valley, our results show that a non-uniform drug distribution does not speed up the evolution of resistance -- indeed, it may actually slow it down. We have verified that these conclusions are also valid for two-dimensional simulations and for more complex drug distributions \cite{us-in-prep}.

Our predictions can be verified experimentally. Recent microfluidic experiments  by Zhang {\em{et al.}} have shown that gradients of the antibiotic ciprofloxacin greatly accelerate the emergence of resistance of the bacterium \emph{E. coli} \cite{austin}.  Although the mutational pathway in this case is not known, our results suggest  that it is likely to be monotonic \footnote{It is important to note that ciprofloxacin increases the mutation rate in {\em{E. coli}}; however we expect this to have a quantitative rather than qualitative effect on our conclusions.}. Furthermore, we predict that repeating the experiments of Zhang {\em{et al.}} using cefotaxime (monotonic pathway \cite{weinreich}) should produce similar results, but  that for streptomycin, which has a fitness valley \cite{schrag1, schrag2}, drug gradients should actually slow down the emergence of resistance. Furthermore, our results may also pave the way to the development of new experimental methods, in which the  characteristics of unknown mutational pathways are deduced by measuring the dependence of the time to resistance on drug concentration and drug gradient.

{\it Note added in revision.} After submission of this manuscript we became aware of related work \cite{HH2}, which addresses a similar model to ours, for the case of the mutational pathway with monotonically increasing MIC.

{\it Acknowledgments.} We thank R. A. Blythe, M. E. Cates, M. R. Evans, W. C. K. Poon, J. Venegas-Ortiz, and P. Warren for helpful discussions. This work was supported by EPSRC under grant number EP/E030173. PG was funded by a DAAD postdoc fellowship, BW by a Leverhulme Trust Early Career Fellowship and RJA by a Royal Society University Research Fellowship and a Royal Society Research Grant.

\end{document}